# ANALYSIS OF S-BAND SUBSTRATE INTEGRATED WAVEGUIDE POWER DIVIDER, CIRCULATOR AND COUPLER

Rahali Bochra[1], Feham Mohammed[1] and Junwu Tao[2]

[1] STIC Laboratory, University of Tlemcen, Tlemcen 13000, Algeria

[2] LAPLACE Laboratory, INP-ENSEEIHT Toulouse, University of Toulouse

*ABSTRACT*

*The Substrate Integrated Waveguide (SIW) technology is a very promising technique with which we can take the advantages of both waveguides and planar transmission lines. Therefore, in [2.1-3] GHz band various microwave components and devices have been designed successfully using Ansoft HFSS software. We then proceeded to the realization of the coupler and then made measurements of the frequency response in the range [2.1-3] GHz using a network analyzer. Thus, results of this modeling are presented, discussed and allow to integrate these devices in planar circuits.*

## 1. INTRODUCTION

Substrate Integrated Waveguide (SIW) technology is one of the most developed platforms so far as it is quite easy to integrate conventional rectangular waveguide into planar circuits. Easy integration and a high quality factor are interesting characteristics of the rectangular waveguide in the technology SIW (RSIW)[1][2]. A large range of SIW components such as bends [3], filters [4], couplers [5], duplexers [6], sixports junction [7], circulators [8] and phase shifters [9] has been studied. Figure 1 illustrate the RSIW which is designed from two rows of periodic metallic posts connected to higher and lower planes mass of dielectric substrate. Figure 2 and 3 shows the similarity of the geometry and the distribution of the electric field between (RSIW) and the equivalent rectangular wave guide [2] [3]. In this paper, [2.1-3] GHz band RSIW components are proposed and optimized. They are essential for many microwave and millimeter-wave integrated circuits and telecommunication systems.

## 2. FUNDAMENTAL RSIW CHARACTERISTICS

For designing (RSIW) (Figure 1) many physical parameters are necessary as d the diameter of holes stems, p the spacing between the holes and $W_{SIW}$ spacing between the two rows of holes. Between the two metal planes of the dielectric substrate two rows of holes are drilled and metalized permitting propagation for all modes $TE_{n0}$ [5]. The lines of current along the lateral walls of the RSIW are vertical, the fundamental $TE_{10}$ mode may be propagated. Electrical performance of RSIW and a conventional rectangular waveguide filled with dielectric of width $W_{eq}$ [4] are similar.

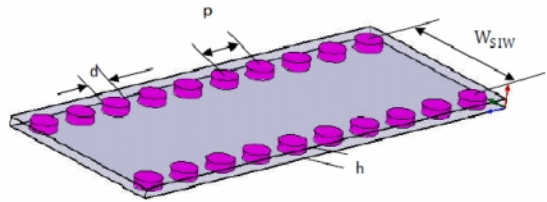

Figure 1. Rectangular wave guide integrated into a substrate RSIW

For obtaining the same characteristics of the fundamental mode propagating in the RSIW (Figure 2) having the same height and the same dielectric, empirical equations [2] were derived in order to determine the width of the equivalent rectangular wave guide

$$W_{eq} = W_{SIW} - \frac{d^2}{0.95\, p} \qquad (1)$$

$$p < \frac{\lambda_0}{2}\, \sqrt{\varepsilon_r} \qquad (2)$$

$$p < 4\, d \qquad (3)$$

$$\lambda_0 = \frac{c}{f}$$

Where $\lambda_0$ is the space wavelength.

Period p must be low for reducing leakage losses between adjoining cylinders. We examined through this study, the RSIW [2.1-3] GHz from a conventional waveguide [10], the feature parameters are outlined in Table 1. We deduce the parameters of RSIW and the equivalent waveguide (Figure 2) Table 1 from the approach cited in [2].

Table 1

| Classic wave guide | Equivalent wave guide | RSIW |
|---|---|---|
| WR340, a=86.36mm, b=43.18mm, $\varepsilon_r = 1$ | h=1.5mm, $\varepsilon_r = 4.3$ $W_{eq}$=42.72mm | h=1.5mm, $\varepsilon_r$ =4.3, d=1mm, p=2mm, $W_{SIW}$=43.25mm |

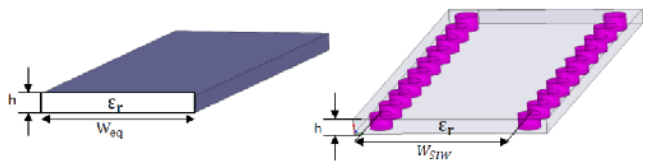

Figure 2. Equivalent rectangular waveguide and RSIW

International Journal of Computer Science, Engineering and Applications (IJCSEA) Vol.4, No.2, April 2014

HFSS tool [11] based on the finite element method (FEM) allows to optimize initial values $W_{SIW}$ given by (1), (2) and (3).It also helps give the scatter diagram and layout of the cartography of the electromagnetic field of the $TE_{10}$ mode. Through Figure 3 we note similarity between electromagnetic field distribution of $TE_{10}$ mode guided in equivalent rectangular waveguide and RSIW.

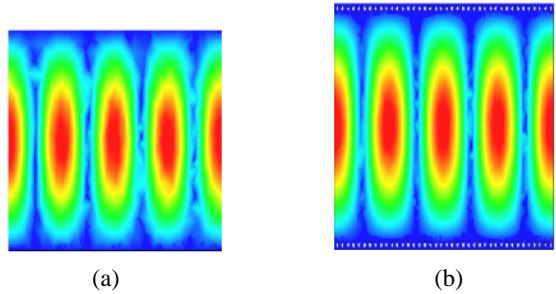

(a)　　　　　　　(b)

Figure 3. Electric field distribution of the $TE_{10}$ mode in the equivalent rectangular waveguide (a) and RSIW (b) at the frequency f = 2.5 GHz

Also Figure 4 shows, between these two waveguides, the coherence of the dispersion characteristics . It is worth noting that the similarity of propagation is valid for all modes $TE_{n0}$.

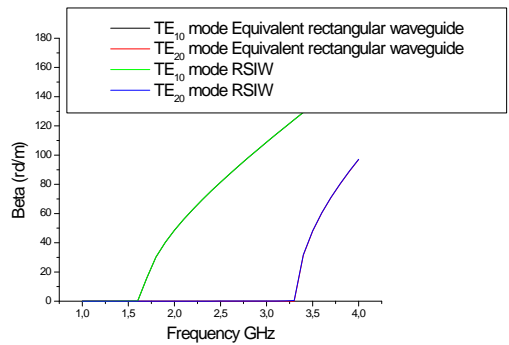

Figure 4. Dispersion characteristics

## 3. RSIW-Microstrip Tapered Transition

For interconnect RSIW to the planar transmission lines the microstrip transition taper [12] is employed .A tapered section is used to match the impedance between a 50 microstrip line in which the dominant mode is quasi-TEM and $TE_{10}$ mode of the RSIW, their electric field distributions are approximate in the profile of the structure.

From several formulas given [13] initial parameters $W_T$ and $L_T$ are determined and optimized with HFSS [11] Table 2. Figure 5, 6 and 7 shows the proposed transitions of coplanar taper of dimensions $L_T$ , $W_T$ to RSIW and mentioned the results of the RSIW simulation without transition and with microstrip line to RSIW.

Table 2

| $L_T$ | 61.5mm |
|---|---|
| $W_T$ | 27.1mm |
| $W_{mst}$ | 3.6mm |
| L | 39.8mm |

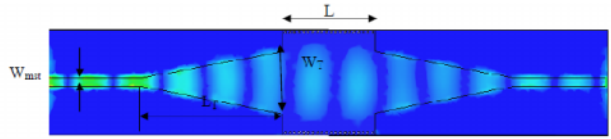

Figure 5. Electric field distribution of $TE_{10}$ mode at f = 2.5 GHz in the matched RSIW

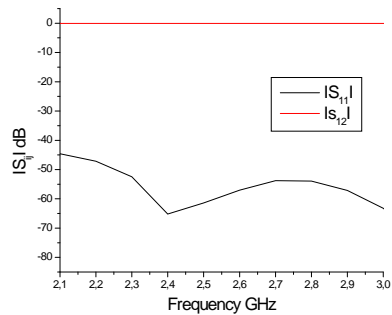

Figure 6. Transmission coefficients S21 and reflection S11 of the RSIW

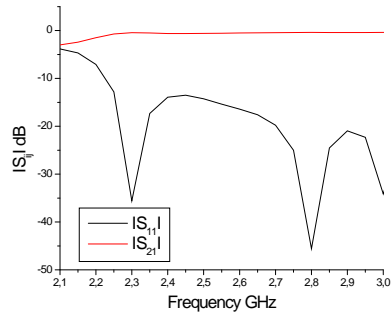

Figure 7. Transmission coefficients S21 and reflection S11
of the matched RSIW with taper

The reflection coefficient $S_{11}$ remains below -15dB over 28.57% of the frequency band and the transmission coefficient $S_{21}$ is around - 0.46 dB across the entire band. Without any mechanical assembly [14] [15], this concept [13] allows the design of a completely integrated planar circuit of microstrip and waveguide on the same substrate.

## 4. DESIGN OF RSIW PASSIVE DEVICES

### 4.1. SIW Circulator

In high power, the circulator based on waveguide technology [8] is still the best solution for protection microwave sources. The RSIW circulator geometry (Figure 8) have three access

International Journal of Computer Science, Engineering and Applications (IJCSEA) Vol.4, No.2, April 2014

separate of 120° from each other, around a central body of ferrite (nickel materials and lithium ferrite) [8] [9]. An incoming wave from port 1, 2 or 3 cannot out by the access 2, 3 or 1, respectively.

The ideal circulator is able to direct the energy to the next access, the third being isolated. Its S matrix (4):

$$[S] = \begin{vmatrix} 0 & 0 & e^{j\varphi} \\ e^{j\varphi} & 0 & 0 \\ 0 & e^{j\varphi} & 0 \end{vmatrix} \quad (4)$$

In this paper the circulator was designed by using cylindrical metal rods (Tables 1 and 2) with L = 20mm and the saturation magnetization of ferrite material is [9] 4 Ms = 5000 Gauss. Its relative dielectric constant is 13.7 and a radius $R_f$ calculated by [8].

$$R_f = \frac{1.84\ c}{\omega_0 \sqrt{\varepsilon_f}} \quad (5)$$

Where c and $\omega_0$ are respectively the velocity of light in the free space and the operation frequency [8]. The ferrite radius and height are $R_f$=6mm $h_f$=1.5mm.

Figure 9 illustrates the distribution of the electric field of the $TE_{10}$ mode circulator RSIW simulated by Ansoft HFSS [11] in [2.1-3] GHz band. The frequency response of RSIW circulator, transmission coefficients $S_{21}$, reflection coefficients $S_{11}$ and isolation coefficients $S_{31}$ are reported through the Figure 10. The reflection loss $S_{11}$ below -15 dB occupy more than 23% of the bandwidth against by the insertion loss $S_{21}$ is in the range of -0.53 dB, while the maximum of the isolation $S_{31}$ is -29.72 dB. At frequency of 2.59 GHz, the two figures 9 and 10 confirm traffic property of the device [14].

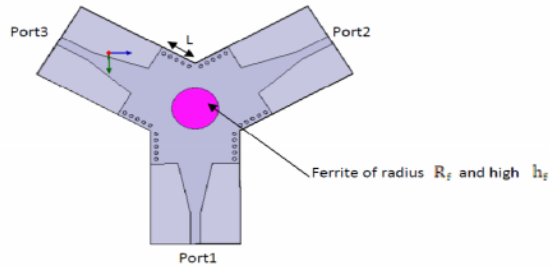

Figure 8. RSIW circulator

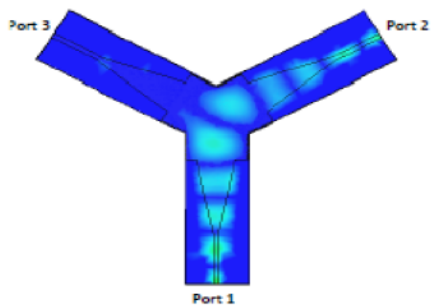

Figure 9. Electric field distribution of the $TE_{10}$ mode of the RSIW circulator at f = 2.67 GHz

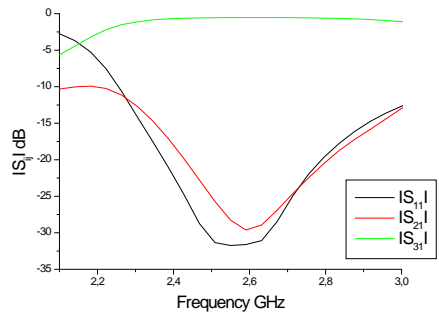

Figure 10. Parameters $S_{ij}$ of RSIW circulator

## 4.2. SIW Power Divider

The power dividers [16] are commonly used to deliver copies of a signal in a system, there are mainly two types T and Y [14] [15]. Our study focuses on the three ports power dividers with equal power division ratio where the half power (-3 dB) of an input signal is provided to each of the two output ports, its S matrix is shown in equation (6)

$$[S] = \begin{vmatrix} S_{11} & S_{12} & S_{13} \\ S_{21} & S_{22} & S_{23} \\ S_{31} & S_{32} & S_{33} \end{vmatrix} \quad (6)$$

The analyzed power divider (Figure 11), designed in the [2.1-3] GHz band, is based on three RSIW of length L=21.3mm connected to form a T. To minimize reflection losses at the input port an inductive metal cylinder of radius r and position xp is added to this power divider. It is generally useful to fix the radius r to the corresponding available practical value of diameter drills, and then change xp to reduce the reflection losses below -15 dB. With added microstrip transition to each port we can integrate power divider directly into a microstrip circuit.

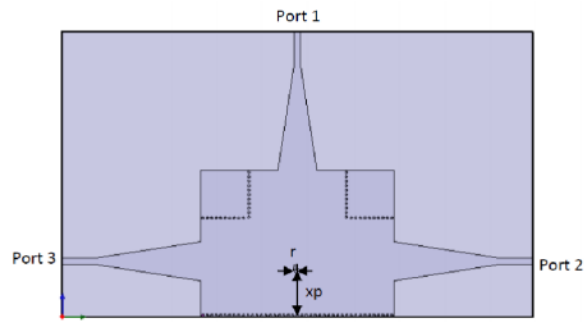

Figure 11. RSIW power divider

The input wave (port 1) is distributed equally into two parts which output to port 2 and port 3 (Figure 12). Figure 13 indicates that $S_{11}$ is less than -15 dB between 2.1 GHz and 2.68 GHz which is more than 24.26 % of the bandwidth. The optimal values of the inductive cylinder are r = 1.2mm, xp=20.57mm. Transmission coefficients $S_{21}$ and $S_{31}$ fluctuate between -3.44dB and -3.61dB being very acceptable levels.

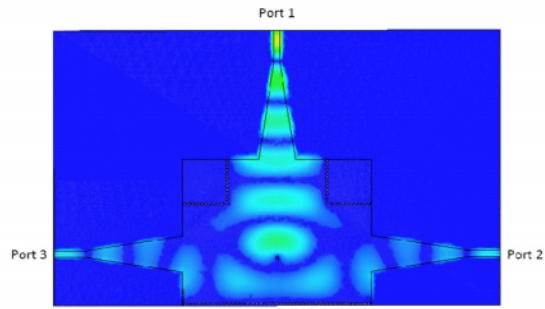

Figure 12. Electric field distribution of the $TE_{10}$ modeat f = 2.4GHz in the RSIW power divider with inductive cylinder

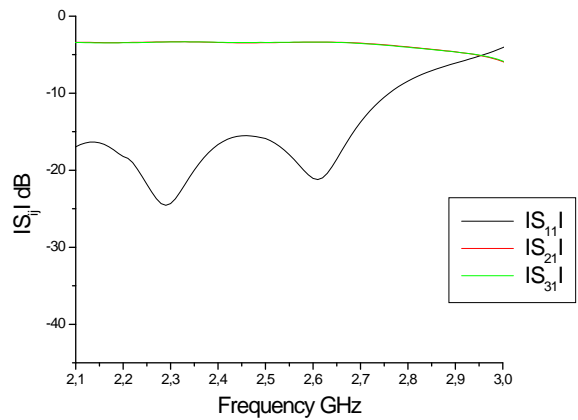

Figure 13. Parameters $S_{ij}$ in the RSIW power divider with inductive cylinder

### 4.3. SIW Coupler

Couplers [5] have been widely used as key components for routing, dividing and combining the signals in the microwave system. Great interest and effort have been directed to the development of different types of directional couplers [17] important element in power dividing/combing networks. The RSIW directional coupler (-3 dB) (Figure 14) is realized by two RSIW with a common wall on which an aperture is utilized to realize the coupling between these two guides. The geometry of the coupler [15] is based on an even/odd mode analysis, where $\beta_1$ and $\beta_2$ are the propagation constants of the $TE_{10}$ and $TE_{20}$ modes, respectively. The phase difference (7)

$$\Delta\varphi = (\beta_1 - \beta_2)W_{ap} \tag{7}$$

To each port the tapered transition between the 50 Ω microstrip line and the RSIW coupler is added integrating this component directly into a microstrip circuit. The S matrix (8):

$$[S] = \frac{1}{\sqrt{2}}\begin{vmatrix} 0 & 1 & j & 0 \\ 1 & 0 & 0 & j \\ j & 0 & 0 & 1 \\ 0 & j & 1 & 0 \end{vmatrix} \tag{8}$$

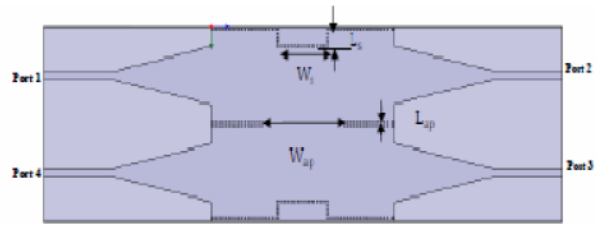

Figure 14 . RSIW coupler

The coupler parameters are finely tuned using HFSS [11] to achieve wide-band performance (Table 3).

Table 3

| | |
|---|---|
| L | 112mm |
| $W_s$ | 30mm |
| $L_s$ | 8mm |
| $W_{ap}$ | 50mm |
| $L_{ap}$ | 3mm |

Figures 15 and 16 show clearly the directional coupler character in the [2.1-3] GHz band where we have the levels of reflection and isolation below -15dB with more than 23.46% of the bandwidth, and which the insertion loss $S_{21}$ and coupling $S_{31}$ fluctuate between -6.50 dB and -7.84 dB. The simulation results prove the good performance of this integrated structure.

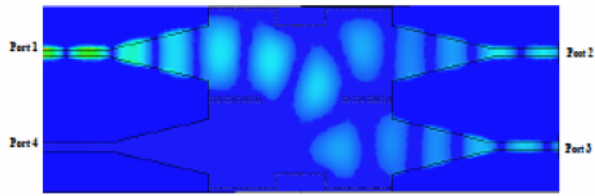

Figure 15. Electric field distribution of TE10 mode for RSIW coupler at f=2.9 GHz

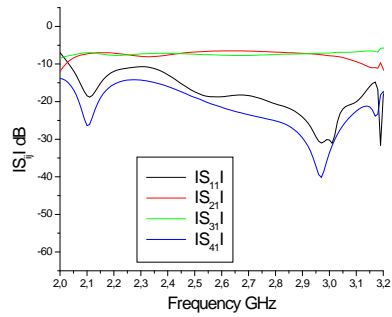

Figure 16. Frequency response of the RSIW directional coupler

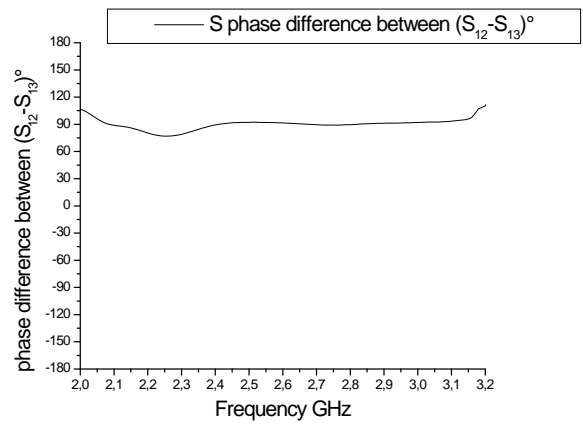

Figure 17. Simulated phase difference

Figure 17 shows the simulated phase difference between two output ports. It can be seen the phase difference is distributed in the range 90.14°~94.66° of within the frequency band of 2.41 to 3.14 GHz.

We then proceeded to the realization of the coupler (figure 18) and then made measurements of the frequency response in the range [2.1-3] GHz using a network analyzer. We were able to take the measurements (Figure 19), modules of the reflection coefficients $S_{11}$, transmission coefficients $S_{12}$, coupling and isolation coefficients $S_{13}$ and $S_{14}$. These values were then plotted on graphs and compared with simulated values.

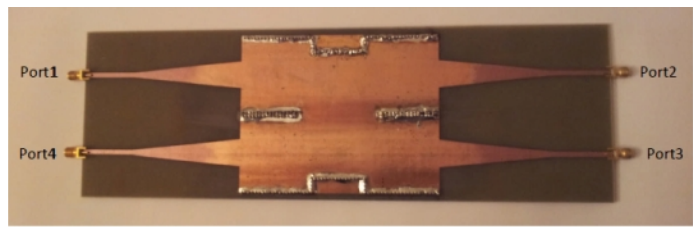

Figure 18. Prototype of SIW coupler

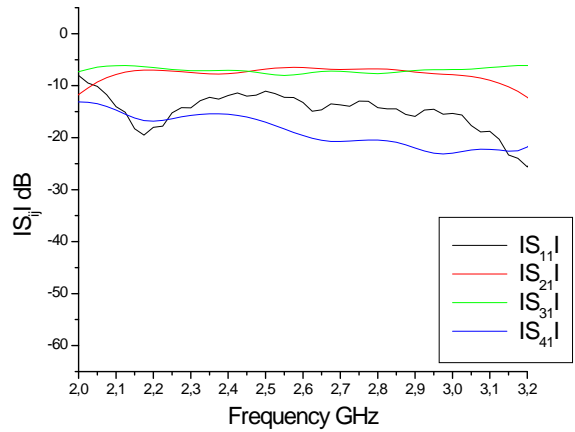

Figure 18. Measured parameters $S_{ij}$ of coupler SIW

The simulated results and the measured results are compared, we note clearly through the Figures 19 (a, b, c, d) a good agreement between simulated and measured values. The results show precisely the directional behaviour of the circuit in a bandwidth of more than 26.11%, corresponding to less than -15 dB reflections. In this band insertion and coupling are respectively on average - 6.46 dB and -7.68 dB. However, the maximum isolation is - 40.02 dB at 2.97GHz. The slight difference between the simulated and measured results from the fact that we did not take into account losses (in the substrate, the metal walls and radiation between the metal rods) during the simulation in HFSS using the finite element method which is rigorous.

We can see that the measured insertion loss is sometimes better than the simulated one. This may be explained by the fact that the substrate loss tangent at 10 GHz given by the manufacturer is used in the simulation.

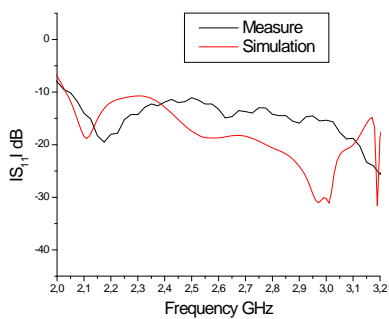
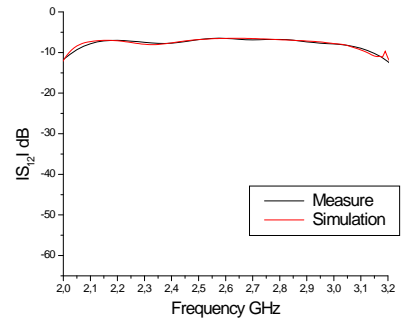

a) réfection coefficients          b) Transmission coefficients

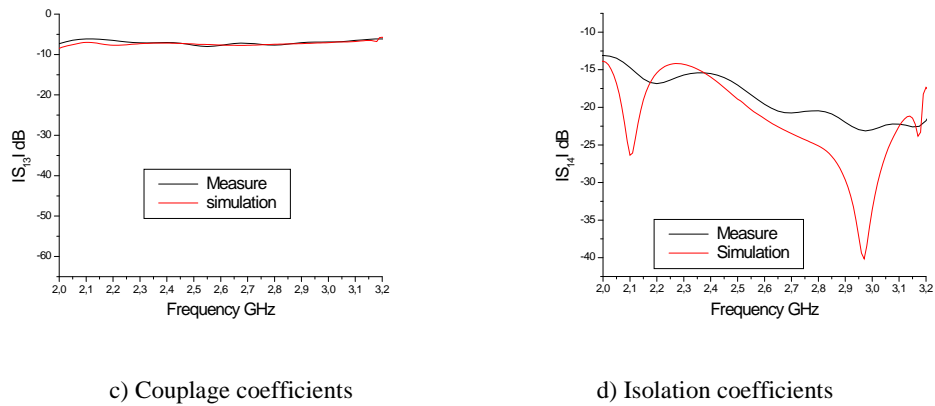

c) Couplage coefficients          d) Isolation coefficients

Figure 19 (a, b, c, d). Comparison of measured coefficients $S_{ij}$ with those simulated

The proposed directional coupler S-band is totally realized by metallic rods in single-layer substrate, and take the advantages of compact size, low-weight and low-cost. It is a favorable choice for designing microwave planar circuits. Prototype of the directional couplers is designed, fabricated using a PCB process and measured with a VNA. Measured results show the directional coupler has good performance in broad operating bandwidths.

## 3. CONCLUSIONS

Through this paper we have investigated a [2.1-3] GHz band substrate integrated waveguide passive components. We have used Ansoft HFSS software to design three RSIW components, the circulator, the power divider and coupler. We then proceeded to the realization of the coupler and then made measurements of the frequency response in the range [2.1-3] GHz using a network analyzer .We note clearly a good agreement between simulated and measured values. The simulation results have shown the good performance of these integrated structures.

## Authors

**Rahali Bouchra** received a Magister degree in physics electronics from, University of Tlemcen (Algeria), in 1986, and the PhD degree in Electromagnetic Modeling of Complex Structures in Microwave Technology SIW in 2013.Since 1982, she has been Assistant Professor and been involved in several research projects at STIC laboratory. Her main area of interest is the simulation of microwave circuits in SIW technology.

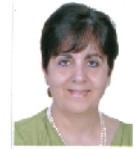

**Feham Mohammed** received the Dr. Eng. degree in optical and microwave communications from the University of Limoges (France) in 1987, and his PhD in Science from the University of Tlemcen (Algeria) in 1996. Since 1987, he has been Assistant Professor and Professor of microwave and communication Engineering. He has served on the Scientific Council and other committees of the Electronics and Telecommunication Departments of the University of Tlemcen. His research interest now is mobile networks and services.

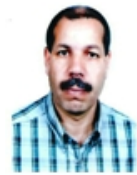